\documentclass[twocolumn,pre,superscriptaddress,showpacs,preprintnumbers,amsmath,amssymb]{revtex4}
\usepackage{graphicx}
\usepackage{dcolumn}
\usepackage{bm}
\usepackage[usenames] {color}
\definecolor{gray}{gray}{0.5} 
\newcommand{\RPA}{{\mbox{\scriptsize RPA}}}

\newcommand{\kb}{k_{\mbox{\scriptsize B}}}

\newcommand{\Li}{\mbox{Li}}

\newcommand{\ave}[1]{\left\langle {#1} \right\rangle}

\begin{document}
\title{
Thermodynamic and
Structural Properties of the High Density Gaussian Core Model}
\author{Atsushi Ikeda}
\affiliation{Institute of Physics, University of Tsukuba, Tennodai 1-1-1, Tsukuba 305-8571, Japan}
\author{Kunimasa Miyazaki}
\affiliation{Institute of Physics, University of Tsukuba, Tennodai 1-1-1, Tsukuba 305-8571, Japan}

\date{\today}
\begin{abstract}
We numerically study thermodynamic and structural properties of the
one-component Gaussian core model (GCM) at very high densities.
The solid-fluid phase boundary is carefully 
determined. 
We find that the density dependence of both the freezing and melting
 temperatures obey the asymptotic relation,  
$\log T_f$, $\log T_m \propto -\rho^{2/3}$, where 
$\rho$ is the number density, which is consistent with Stillinger's conjecture.
Thermodynamic quantities such as the energy and pressure and the
 structural functions such as the static structure factor are also
 investigated in the fluid phase for a wide range of temperature above the
 phase boundary. 
We compare the numerical results with the prediction of the
liquid theory with the random phase approximation (RPA). 
At high temperatures, the results are in almost perfect agreement with
 RPA for a wide range of density, as it has been already shown in
 the previous studies. 
In the low temperature regime close to the phase boundary line, 
although RPA fails to describe the structure factors and the radial distribution 
functions at the length scales of the interparticle distance, 
it successfully predicts their behaviors at shorter length scales.
RPA also predicts thermodynamic quantities such as the energy, pressure, 
and the temperature at which the thermal expansion coefficient
becomes negative, almost perfectly. 
Striking ability of RPA to predict thermodynamic quantities 
even at high densities and low temperatures is understood in terms of 
the decoupling of the length scales which dictate thermodynamic quantities
from the interparticle distance which dominates the peak structures of
the static structure factor due to the softness of the Gaussian core potential. 

\end{abstract}
 \maketitle

\section{Introduction}

Complex fluids such as colloidal suspensions and emulsions
are often regarded as macroscopic models of atomic or molecular systems. 
They are ideal benches to test liquid theories developed to describe thermodynamic, dynamic, and
structural properties of atomic and molecular liquids~\cite{Hansen2006}. 
It is not only because the size of constituent unit of complex fluids
are much larger than atomic counterpart but also because their
interparticle interactions can be tailor-made and tuned relatively easily. 
While the pair interactions of atomic systems are exclusively 
characterized by short-ranged and strong repulsions with weak and longer-ranged
attractions, leading to 
the typical phase diagram demarcating gas, liquid, and crystalline
phases~\cite{Hansen2006}, those for the complex fluids are far more
diverse.
This diversity leads to very rich and often counter-intuitive macroscopic behaviors~\cite{Likos2001a}.  
Amongst those interactions, the {\it ultra-soft} potential have
attracted particular attention recently in the soft-condensed matter community~\cite{Likos2006b,
Stillinger,Stillinger1997,Lang2000,Louis2000a,Prestipino2005,
Mladek2006b,Mausbach2006,Zachary2008,Krekelberg2009c,Pond2009,Krekelberg2009d,Shall2010,Pond2011,
Marquest1989,Likos1998,Likos2001b,Mladek2006a,Zhang2010,
Watzlawek1999,Foffi2003b,Zaccarelli2005,Mayer2008,
Pamies2009,Berthier2009f,Berthier2010e}.
The ultra-soft potentials are isotropic repulsive potential 
characterized by weak and bounded repulsion at short distance 
and the mild repulsive tails whose steepness is much smaller than
the typical atomic potential. 
These potentials are realized in many complex fluids such as
star-polymers~\cite{Watzlawek1999,Foffi2003b,Zaccarelli2005,Mayer2008}, 
dendrimers~\cite{Likos2006b,Gotze2005,Mladek2008}, 
and the polymers in good solvent~\cite{Louis2000a,Kruger1989,
Dautenhahn1994,Louis2000b,Bolhuis2001}. 
Thermodynamic phase diagrams of the ultra-soft particle systems have
very distinct and exotic properties such as the re-melting from solid to fluid phase at high densities, 
the re-entrant peak at the intermediate densities~\cite{Stillinger,Lang2000,Prestipino2005,
Zachary2008,Zhang2010,Watzlawek1999,Pamies2009}, 
negative thermal expansion coefficient~\cite{Stillinger1997,Mausbach2006}, and the cascades of
the various crystalline phases at very high densities~\cite{Watzlawek1999,Pamies2009}. 
The Gaussian core model (GCM) is one
of the simplest examples of the ultra-soft potential systems. 
GCM consists of the point particles interacting with a Gaussian shaped repulsive potential;
\begin{eqnarray}
v(r) = \epsilon \exp[-(r/\sigma)^2], 
\end{eqnarray}
where $r$ is the interparticle separation, 
$\epsilon$ and $\sigma$ are the parameters which characterize the energy
and length scales, respectively. 
GCM was first introduced by Stillinger~\cite{Stillinger} and has been studied 
by many groups~\cite{Stillinger1997,Lang2000,Louis2000a,Prestipino2005,
Mladek2006b,Mausbach2006,Zachary2008,Krekelberg2009c,Pond2009,Krekelberg2009d,Shall2010,Pond2011}.  
Despite of its simple form of the pair potential, 
GCM exhibits many typical thermodynamic behaviors of the ultra-soft
particles.
According to thermodynamic phase diagram obtained by numerical
simulations~\cite{Stillinger,Prestipino2005},   
GCM basically behaves like hard spheres at very low densities
and temperatures;
the crystalline structure in the solid phase is fcc and the freezing/melting
temperatures sharply increase with density. 
However, as the density increases further, the freezing temperature 
reaches a maximal value and beyond this point 
it changes to a decreasing function of the density.
This re-entrance takes place at $\rho\sigma^{3}\approx 0.25$, where 
$\rho$ is the number density.
Concomitantly, the crystalline structure changes from fcc to bcc.
Recently, thermodynamic and transport anomalies of the fluid phase in the vicinity of the reentrant
peaks are investigated ~\cite{Mausbach2006,Krekelberg2009c,Pond2009,Krekelberg2009d,Shall2010,Pond2011}. 
Microscopic and structural properties such as the static structure factor
in the fluid phase are also reported and documented~\cite{Lang2000,Louis2000a,Mladek2006b,Zachary2008}.   
These studies revealed that, as the density increases beyond the
reentrant peak but at a fixed temperature, thermodynamic and
structural properties of GCM becomes more ideal-gas-like, signaled by 
the lowering of the peak of the structure factors and the better
agreement with simple approximations such as the random phase
approximation (RPA).
Most of studies in the past, however, have focused on the densities not
far from the reentrant peak or the relatively high temperatures.  
Less attention has been paid for the high density and low
temperature regimes, especially in the vicinity of the solid-fluid phase
boundary. 
Near the phase boundary line, the thermodynamic and structural properties 
are expected to be highly non-trivial even at the high density limit. 
Based on the duality argument of the ground state of GCM in the
reciprocal space, Stillinger has conjectured that the freezing and melting
temperatures, $T_f$ and $T_m$, are given by an asymptotic form; 
\begin{eqnarray}
\log T_f, ~\log T_m \propto -\rho^{2/3}  
\label{st}
\end{eqnarray}
in the high density limit~\cite{Stillinger}. 
However, this conjecture has not been confirmed  numerically. 
Recently, we studied the nucleation and glassy dynamics of the one-component
GCM in the supercooled state at the unprecedentedly high densities,
$0.5 \leq \rho\sigma^3 \leq 2$~\cite{Ikeda2011}. 
It was found that the crystal nucleation rate decreases drastically as 
the density increases and concomitantly dynamics of the constituent particles
becomes very sluggish. 
The density time correlation function exhibits typical behavior of the
supercooled liquids near the
glass transition point, such as the two-step and
non-exponential structural relaxation. 
The relaxation time steeply increases as the temperature is lowered at a
fixed density.  
Surprisingly, these are well described by the mode-coupling theory,
implying that the high density and one-component GCM is more amenable to the
mean-field picture of the glass transition than
other typical glass formers. 
These observations call for more detailed analysis of the high density GCM
at the low temperature regime.
Especially, it is tempting to consider GCM in the high density limit 
as the ideal and clean model system to study the glass
transition. 
Thermodynamic and structural characterization are prerequisites for
dynamical study~\cite{Ikeda2011,Ikeda_II} 
but the detailed study is lacking. 

In this work, we numerically investigate thermodynamic and structural
properties of the one-component GCM up to the density $\rho\sigma^3 = 2.4$. 
We determine the solid-fluid phase boundary and show that the
Stillinger's scaling, Eq.~(\ref{st}), holds at $\rho\sigma^3 \gtrsim 1.2$. 
Thermodynamic and microscopic structural
properties of GCM are also analyzed carefully
over a wide range of temperature and density. 
The potential energy, pressure, thermal expansion coefficient, and the
static structure factors are evaluated 
and compared with the prediction of 
the liquid state theory.
Surprisingly good agreement with the random phase approximation (RPA) is
found for thermodynamic quantities for a wide range of temperature,
including the low temperature regimes where 
the same approximation poorly describes the static structure factor and
radial distribution function. 
This counterintuitive observation can be attributed to the ultra-soft
nature of GCM for which the microscopic structure near the first shell of
the system decouples with the macroscopic properties. 

This paper is organized as follows. 
In Section II, technical details of simulations and the method to
compute the phase boundary are discussed. 
In Section III, we  present simulation results for the phase diagram,
thermodynamic quantities, and structural functions. 
We compare the simulation results in the fluid phase with RPA predictions in
Section IV. 
We summarize and discuss the results in Sec. V.

\section{Simulation Method}

\subsection{MD and MC simulation}

Thermodynamic state of GCM is fully characterized by the density $\rho$ and temperature $T$. 
In this work, we focus on the density and temperature range of 
$0.3 < \rho^{\ast} < 2.4$ and $10^{-6} < T^{\ast} < 1$, where 
$\rho^{\ast}\equiv \rho\sigma^{3}$ and $T^{\ast} \equiv \kb T/\epsilon$.  
In order to analyze thermodynamic properties and determine the solid-fluid phase boundary, 
molecular dynamics (MD) simulations with Nos{\'e} thermostat are carried out under the periodic boundary conditions. 
To integrate the equations of motion, we use a reversible algorithm similar
to the Velocity-Verlet method~\cite{Frenkel2001} with time steps of $0.1\tau$,  
which is sufficiently short to preserve the Nos{\'e} Hamiltonian. 
Here $\tau = \sqrt{m\sigma^2/\epsilon}$ is the time unit, where $m$ is
the mass of a particle. 
For evaluation of the free energy of the reference state (see below), Monte Carlo (MC) simulation is used.
In a trial MC move, 
the maximum displacement of a particle is adjusted to keep the
acceptance ratio about 50 \%.  
In both simulations, 
the total number of particles is $N=3456$. 
This is twice the cube of an integer (in this case 12), 
a natural choice for the bcc crystal in a cubic simulation box.  
The cutoff length of the potential is taken as $5\sigma$. 
The pair potential at the cutoff length is $1.4 \times 10^{-11} \epsilon$  
which is much smaller than the typical kinetic energy at the lowest
temperature studied in this work. 

\subsection{Evaluation of the free energy}

The chemical potential as a function of the temperature and pressure, $\mu(T,P)$, 
is required in order to determine the phase boundary.  
We evaluate it using the free energy $f(T,\rho)$ and pressure $P(T,\rho)$. 

We calculate the free energy using the thermodynamic integration method combined with 
the particle insertion method~\cite{Widom1963} for the fluid phase and the Frenkel-Ladd
method ~\cite{Frenkel1984} for the crystalline phase. 
This procedure is the same as the one employed by Prestipino {\it et al.}~\cite{Prestipino2005}
for lower densities. 
The free energy of the system is the sum of the ideal $f_{id}$ and excess part $f_{ex}$. 
$f_{id}$ is given by 
$f_{id}(T,\rho) = \kb T (\log \Lambda^3 \rho \sigma^3 - 1)$, 
where $\Lambda = \sqrt{2\pi \hbar^2/m\kb T}$ is the de Broglie thermal wave length.
According to the thermodynamic integration scheme, $f_{ex}(T,\rho)$ 
can be evaluated by integrating over the energy and pressure 
from the reference state point $(T_0,\rho_0)$  to the target state point
$(T,\rho)$ using the following equations.
\begin{equation}
\begin{aligned}
\frac{f_{ex}(T,\rho_0)}{T} 
&= \frac{f_{ex}(T_0,\rho_0)}{T_0} + \int^{T}_{T_0}\!\!\! dT' \ \frac{u(T',\rho_0)}{T'^2},
\\
\frac{f_{ex}(T,\rho)}{T} 
&= \frac{f_{ex}(T,\rho_0)}{T} + \int^{\rho}_{\rho_0}\!\!\! d\rho' \!\!\ 
\left\{
\frac{P(T,\rho')}{\rho'^2 T} - \frac{1}{\rho'}
\right\},
\end{aligned}
\label{ti2} 
\end{equation}
where $u$ is the potential energy per particle. 
For the fluid phase, the reference free energy $f_{ex}(T_0,\rho_0)$ is
calculated by the particle insertion method~\cite{Frenkel2001,Widom1963}
and the pressure is evaluated from the virial equation.
In this method, the free energy is expressed in terms of the
energy cost to insert one particle into the system as 
\begin{eqnarray}
\frac{f_{ex}(T_0,\rho_0)}{T_0} = - \kb\mbox{log} \ave{\exp
 \left(-\frac{E_{in}}{\kb T_0}\right)} -
 \frac{P_0}{\rho_0 T_0} + 1,  
\label{particleinsertion}
\end{eqnarray}
where $E_{in}$ is the interaction energy of an inserted particle with other particles in the system. 
The average should be taken over the ensemble of randomly inserted
particles. 
For the crystalline phase, on the other hand, the reference free energy is computed using the Frenkel-Ladd 
method~\cite{Frenkel2001,Frenkel1984} 
which is a different kind of thermodynamic integration technique.  
In this method, we consider a hybrid Hamiltonian $\tilde{V}(\lambda)$, which interpolates
between the Hamiltonian of the original system $V$ and that of the Einstein crystal $V_{ein}$ 
as $\tilde{V}(\lambda) = V + (1-\lambda) V_{ein}$, where $\lambda$ is the switching parameter.
The free energy of the original system can be computed by the following
equation, which is the integral over $\lambda$ of the Hamiltonian of the
hybrid system evaluated from the simulation, 
\begin{eqnarray} 
f_{ex} = f_{ex,ein} + \frac{1}{N}\int^{1}_0 d\lambda \ \ave{V - V_{ein}}_{\lambda}, 
\label{fl}
\end{eqnarray} 
where $\ave{\cdots}_{\lambda}$ is the ensemble average under a hybrid Hamiltonian $\tilde{V}(\lambda)$ 
and $f_{ex,ein}$ is the excess part of the free energy of the Einstein crystal. 
We choose $(T_0^{\ast}, \rho_0^{\ast})$ $=$(0.1, 0.01) and
($T_0^{\ast}$, $\rho_0^{\ast}$) $=$
(0.0794, 0.28) as the reference states for the fluid and crystalline phases,
respectively.  
The ensemble averages in Eqs.~(\ref{particleinsertion}) and (\ref{fl})
are evaluated using the MC simulations. 
The integration over $\lambda$ in Eq.~(\ref{fl}) is calculated by
slicing $\lambda$ to the grids of the width of 0.05 and evaluating
the value of the integrand at each grid point from independent MC simulations.
Likewise, the integral over the isothermal and isochoric pathways in Eq.~(\ref{ti2})
is computed by slicing the pathways into many grid points. 
The energy and pressure at each grid point are computed using the MD
simulations. 
The free energy for both the fluid and crystalline phases are obtained by 
combining these data points and the reference free energy. 
In order to determine the solid-fluid phase boundary over the density
range of $0.3<\rho^{\ast}<2.4$ with satisfactory accuracies, 
more than 800 grid points were necessary.

\section{Simulation results}

\subsection{Phase diagram}

In his pioneering work, Stillinger has conjectured that the solid-fluid
phase boundary is asymptotically given by Eq.~(\ref{st}) in the high
density limit~\cite{Stillinger}.  
This conjecture is based on analysis of the density dependence of the potential energy
of various crystalline structures (bcc, fcc, etc) at $T=0$. 
According to his analysis, the potential energy is expressed as 
\begin{eqnarray}
u = - \frac{\epsilon}{2} + \frac{\pi^{3/2} \rho \sigma^3 \epsilon}{2} \left\{  1 +
 A \exp(- K \rho^{2/3}) + \cdots \right\}, 
\label{st1} 
\end{eqnarray}
where $A$ and $K$ are constants which depend on the crystalline structure. 
Stillinger argued that this density dependence remains qualitatively
unchanged at finite temperatures. 
Since the ordered structure of the crystalline phase is responsible for the term 
$\exp(- K \rho^{2/3})$ in Eq.~(\ref{st1}), the energy difference between
the crystalline and fluid phase at the phase 
boundary should be also proportional to this term. 
This argument leads us to a conjecture that the melting/freezing temperature is proportional
to this factor, which is Eq.~(\ref{st}). 
Here we  verify this argument numerically. 

\begin{figure}[t]
\begin{center}
\includegraphics[width=0.95\columnwidth]{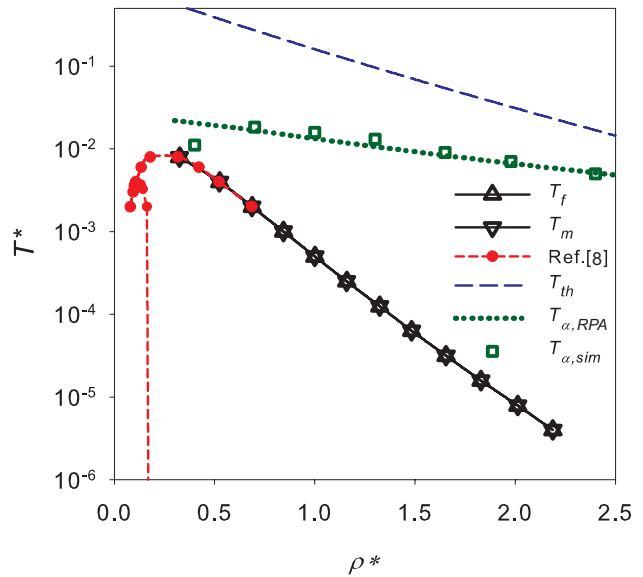}
\caption{
The freezing ($T_f^{\ast}$) and melting ($T_m^{\ast}$) temperatures of GCM 
as a function of density (open up/down triangles). 
The result of Prestipino {\it et al.} is also plotted (filled
 circles with short-dashed line)~\cite{Prestipino2005}.  
The long-dashed line is the threshold temperatures, $T_{th}$, above which RPA gives 
a reasonable description of the system (see Sec. IV).  
Open squares and dotted line indicate the temperature below which the thermal
 expansion coefficient becomes negative, $T_{\alpha}$, obtained from simulation and RPA, respectively.  
The short-dashed line at 
$\rho^{\ast}\approx 
0.15
$ demarcates the fcc
 (left) and bcc (right) crystalline phases.
}
\vspace*{-0.3cm}
\label{spd}
\end{center}
\end{figure}
\begin{figure}[h]
\begin{center}
\includegraphics[width=0.7\columnwidth]{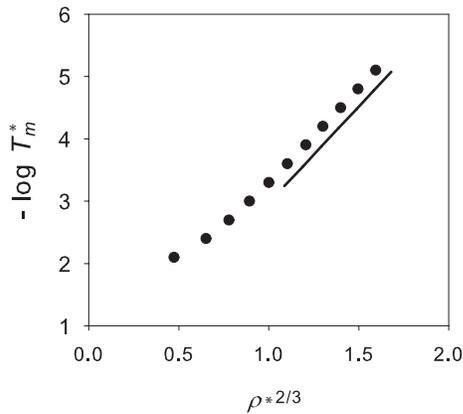}
\caption{
$-\log T^{\ast}_m$ versus $\rho^{\ast 2/3}$. 
The solid straight line is a guide for the eyes. 
}
\vspace*{-0.3cm}
\label{stillinger}
\end{center}
\end{figure}

Figure~\ref{spd} presents the phase diagram of GCM obtained from our simulation. 
Both the freezing and melting temperature $T_f^{\ast}$ and $T_m^{\ast}$ are shown in
this figure, but their values are very close to each other and
indistinguishable in the scale of the figure.  
As expected, the melting and freezing temperatures dramatically
decrease as the density increases, down to $T^{\ast} \approx 10^{-6}$ at the
highest density we studied.
The phase boundary for $\rho^{\ast} \lesssim 0.7$ obtained by Prestipino
{\it et al.}~\cite{Prestipino2005} is also plotted, in order to confirm that  
the present result perfectly matches with theirs for the density
window where both results are available. 
The crystalline structure at high densities is bcc, 
as has been verified by the direct MD simulation of the nucleation~\cite{Ikeda2011,Ikeda_II}. 
In order to verify the scaling relation, Eq.~(\ref{st}), we plot the
logarithm of the melting temperature as a function of $\rho^{2/3}$ in Fig.~\ref{stillinger}. 
One observes that the result rides on the scaling function at $\rho^{\ast} \gtrsim 1.2$.

\subsection{Potential energy and pressure}

\begin{figure}[h]
\begin{center}
\includegraphics[width=0.7\columnwidth]{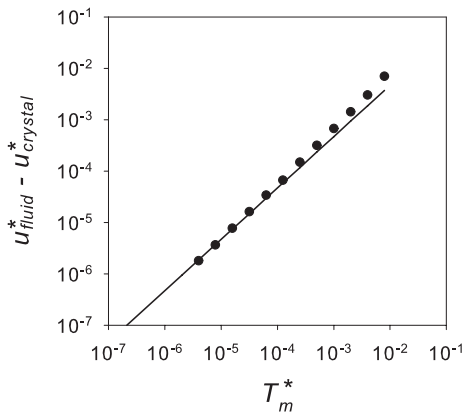}
\caption{
The potential energy difference between the crystalline and fluid phases
 against the melting temperatures with the fit by a straight line of
 the slope 1 (solid line). 
}
\vspace*{-0.3cm}
\label{du}
\end{center}
\end{figure}

The first order transition from the crystalline to fluid phase is
accompanied with the discontinuous change in the structural order and the potential energy. 
We show the temperature dependence of the potential energy difference
between the two phases 
$u^{\ast}_{fluid}(T_f,\rho) -u^{\ast}_{crystal}(T_f,\rho)$ as a function
of $T_m$ in Fig.~\ref{du},  
where $u^{\ast}= u/\epsilon$ is the dimensionless potential energy.
This figure shows that the energy difference is proportional to the
melting temperature at the low temperatures/high densities, 
verifying the assumption which Stillinger has employed to conjecture
Eq.~(\ref{st}). 
The entropy difference between the two phases can be estimated from 
this energy difference by 
$s_{fluid}(T_f,\rho) - s_{crystal}(T_f,\rho) = \left\{u_{fluid}(T_f,\rho) - u_{crystal}(T_f,\rho)\right\}/T_f$.  
From the result of Fig.~\ref{du}, the entropy
difference at the low temperature/high density limit can be estimated as
\begin{eqnarray}
s_{fluid}(T_f,\rho) - s_{crystal}(T_f,\rho) \sim 0.45 \kb. 
\end{eqnarray}
This value should be compared with the results at lower densities in the earlier work;
$0.81\kb$ at $\rho^{\ast} = 0.4$ and $0.54\kb$ at $\rho^{\ast} = 1.0$~\cite{Stillinger}. 

\begin{figure}[h]
\begin{center}
\includegraphics[width=0.95\columnwidth]{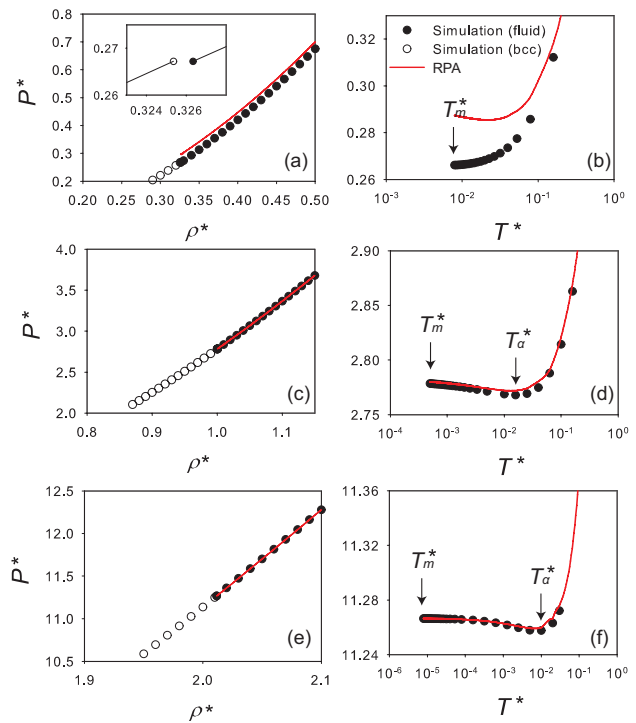}
\caption{
The equation of state of GCM. 
Filled and open circles are the results for fluid and crystalline (bcc) phases, respectively. 
Solid lines are the results from RPA (see Sec. IV).  
Left panels are $P$-$\rho$  plots at $T^{\ast}= $ 
$7.94\times 10^{-3}$ (a), $5.0\times 10^{-4}$ (c), and $7.9\times
 10^{-6}$ (e). 
The inset in (a) is the  closeup around the freezing transition density. 
Right panels are $P$-$T$ plots at $\rho^{\ast}=$ 0.33 (b), 0.99 (d), and
 2.01 (f). 
$T^{\ast}_m$ and $T^{\ast}_{\alpha}$ (see text) are indicated by arrows.
}
\vspace*{-0.3cm}
\label{prhot}
\end{center}
\end{figure}

Next we focus on the equation of state (EOS) of the high density GCM, {\sl i.e.}, the pressure
as a function of the density and temperature. 
The dimensionless pressure $P^{\ast} \equiv P\sigma^3/\epsilon$ is plotted in Figure~\ref{prhot}. 
In Figures~\ref{prhot} (a), (c), and (e), we plot the isothermal cut of
EOS and the isochoric cut in Figures~\ref{prhot} (b), (d), and (f). 
The parameters $T^{\ast}$ and $\rho^{\ast}$ for each adjacent figures have
been chosen so as for them to share the common freezing points; 
Fig.~\ref{prhot} (a) and (b) share the freezing point 
($T_{f}^{\ast}, \rho_{f}^{\ast})=(7.94\times 10^{-3}, 0.33$), (c) and (d) share 
($T_{f}^{\ast}, \rho_{f}^{\ast})=(5.0\times 10^{-4}, 1.00$), and (e) and (f) share
($T_{f}^{\ast}, \rho_{f}^{\ast})=(7.9\times 10^{-6}, 2.01$).  
Figure~\ref{prhot} (a) shows that the melting of the bcc crystalline phase
(white circles) to the fluid phase (filled circles) takes place at
at $\rho^{\ast}\approx 0.33$.
The inset shows the narrow coexistence region around the transition density, 
at which the pressure becomes constant. 
Similar behaviors of the first order transition are also observed in
Figs.~\ref{prhot} (c) and (e) but at much higher densities.

Figure~\ref{prhot} (b) shows the equation of state at
$\rho^{\ast}=0.33$ over the temperature range of $1.0\times 10^{-3}<T^{\ast}<1.0$.  
At this relatively low density, the pressure is a monotonically increasing
function of the temperature, which is usual behavior of ordinary fluids. 
However, at higher densities, as shown in Fig.~\ref{prhot} (d) and (f), 
there exists the temperature regime in which the pressure becomes a decreasing function of temperature. 
In this regime, the thermal expansion coefficient $\alpha=V^{-1}(\partial V/ \partial T)_P$ becomes negative. 
We determine the threshold temperature $T_{\alpha}$ at which $\alpha$
changes its sign by fitting the pressure by a smooth  polynomial function 
and plot $T_{\alpha}$ in Fig.~\ref{spd} (open squares). 
At low densities, $T_{\alpha}$ is located in the vicinity of the phase boundary. 
With increasing density, however, the difference between $T_{\alpha}$ and $T_m$ 
increases monotonically.
Although the existence of the anomalous negative thermal expansion
coefficient of GCM has been reported in the literatures~\cite{Stillinger,Stillinger1997},
its high density behavior has not been explored. 
We shall discuss this result and its asymptotic behavior of $T_{\alpha}$ at high
densities in the following section.

\subsection{Static structure factor}

\begin{figure}[h]
\begin{center}
\includegraphics[width=0.95\columnwidth]{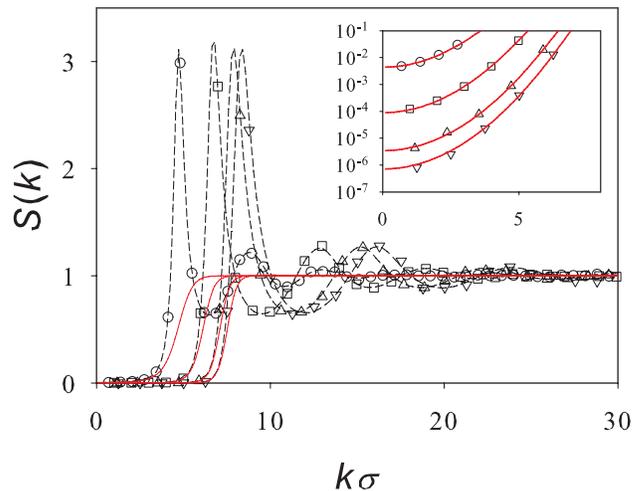}
\caption{
The static structure factors in the fluid phase at the freezing
 temperatures for 
$\rho^{\ast}=0.33$ ($\bigcirc$), $1.00$ ($\square$), $1.65$ ($\bigtriangleup$), and $2.01$ ($\bigtriangledown$).
The dotted lines are the results obtained by the Fourier transformation of $g(r)$. 
The solid lines are the results of RPA (see Sec. IV).  
The inset is the semi-log plot of the main figure for $S(k) \leq 10^{-1}$ at low $k$'s. 
}
\vspace*{-0.3cm}
\label{sk}
\end{center}
\end{figure}

In Figure~\ref{sk}, we show the static structure factors $S(k)$ in the
fluid phase just above the freezing temperatures. 
$S(k)$ obtained from the Fourier transformation of the radial
distribution function $g(r)$ is also shown. 
At all densities, $S(k)$'s exhibit sharp peaks 
similar to those of the ordinary simple fluids, 
such as the hard sphere and Lennard-Jones fluid, near
the freezing temperatures. 
The peak position $k_{\max}$ shifts to high $k$'s as the density increases, 
$k_{\max} \propto \rho^{1/3}$, as one should expects at high densities.
Note that, however, the height of the first peak is about 3.1 at all
densities, which is slightly higher than the universal value 2.85 for 
the ordinary fluids which is known as Hansen-Verlet
criterion~\cite{Hansen2006,Hansen1969}.  

\section{Random phase approximation analysis}

\subsection{Random phase approximation}

It is known that the thermodynamics and microscopic structure of 
the GCM fluid in high densities and temperatures are described by the random phase approximation (RPA)
remarkably well~\cite{Lang2000,Louis2000a}.      
RPA is one of the approximation schemes of the liquid theory 
and a kind of the mean-field theory for the thermodynamics and structure
of liquids~\cite{Hansen2006}. 
In this section, we discuss how this approximation works at much lower
temperatures. 

It is convenient to divide thermodynamic quantities 
into uniform and fluctuation parts as
follows.
The potential energy can be represented as  
\begin{eqnarray}
u &=& - \frac{\epsilon}{2} + \frac{\pi^{3/2} \rho\sigma^3 \epsilon}{2} + 
\Delta u, 
\label{ueq} 
\end{eqnarray}
where the first two terms are the uniform part and $\Delta u$ is the
fluctuation part. 
$\Delta u$ can be expressed as
\begin{eqnarray}
\Delta u &=& \frac{1}{4\pi^2} \int^{\infty}_0 k^2 \tilde{v}(k) S(k) \ dk, 
\label{du1}
\end{eqnarray}
where 
\begin{equation}
\tilde{v}(k) = \pi^{3/2} \epsilon \sigma^3 \exp(-k^2 \sigma^2/4) 
\end{equation}
is the reciprocal expression of $v(r)$. 
Likewise, the pressure can be written, using the virial equation, as 
\begin{eqnarray}
P &=& \kb T \rho + \frac{\pi^{3/2} \rho^2 \sigma^3 \epsilon}{2} + \Delta P,
\end{eqnarray}
where the first two terms are the uniform part and the third is the 
fluctuation part which can be written as
\begin{eqnarray}
\Delta P &=& \frac{\rho}{4\pi^2} \int^{\infty}_0 \bigl( k^2 - k^4\sigma^2/6 \bigr) \tilde{v}(k) S(k) \ dk. \label{dz}
\end{eqnarray}
In RPA, the direct correlation function of the system is approximated 
as $c_{\RPA}(r) = -\beta v(r)$, where $\beta = 1/\kb T$. 
This approximation makes it possible to express
various static quantities in simple and analytic forms. 
The static structure factor can be expressed as 
\begin{eqnarray}
S_{\RPA}(k) = \frac{1}{1+\rho \beta \tilde{v}(k)}. \label{mf}
\end{eqnarray}
The fluctuation parts of the potential energy and pressure are expressed as~\cite{Louis2000a}: 
\begin{equation}
\begin{aligned}
\Delta u_{\RPA} 
&= - \frac{\epsilon}{2\gamma} \Li_{3/2} (- \gamma ), \\
\Delta P_{\RPA} 
&= - \frac{\rho \epsilon}{2\gamma} \left\{ \Li_{3/2} (- \gamma ) - \Li_{5/2}(- \gamma ) \right\}, 
\label{upRPA}
\end{aligned}
\end{equation}
where $\gamma = \pi^{3/2}\rho \sigma^3 \epsilon/\kb T$ is a dimensionless coupling parameter and 
$\Li_{\nu}(x)$ is the $\nu$-th polylogarithm function~\cite{Louis2000a,Mladek2006b}. 
Furthermore, the radial distribution function at $r=0$ can be expressed
analytically as
\begin{eqnarray}
g_{\RPA}(r=0) = 1 + \frac{1}{\pi^{3/2}\rho\sigma^3} \Li_{3/2} (-\gamma).
 \label{gRPA}
\end{eqnarray}
The second term on the right hand side of this expression is negative
for arbitrary densities and temperatures. 
At a very low temperature, the modulus of this term becomes larger than
the first, leading to an unphysical negative $g(r=0)$. 
We refer to this temperature as the threshold temperature $T_{th}$.
We plot $T_{th}$ in Fig.~\ref{spd} (long-dashed line).
At high densities, $T_{th}$ can be expressed analytically as 
\begin{eqnarray}
\frac{\kb T_{th}}{\epsilon} = \pi^{3/2} \rho \sigma^3
\exp\left[- \Bigl( \frac{3\pi^2}{4} \rho\sigma^3 \Bigr)^{2/3} \right], 
 \label{th}
\end{eqnarray}
which is obtained by the asymptotic expansion of polylogarithm
function of Eq.~(\ref{gRPA}) (see Appendix).
Interestingly, the threshold temperature follows the same asymptotic
scaling law $\log T_{th} \propto - \rho^{2/3}$ as the melting and
freezing temperatures, Eq.~(\ref{st}). 
Note that this asymptotic expression is very accurate down to moderate densities $\rho^{\ast} \sim 1$.   

\subsection{High temperature regime}

\begin{table}[b]
\begin{center}
\caption{
RPA results compared with simulation results for the
 fluctuation parts of the potential energy and pressure
at the threshold temperatures for various densities.
}
\label{table1}
\begin{tabular}{l llll}  \hline \hline
$\rho$ &  0.10 &  0.33 & 1.00 &  3.00 \\ 
${T_{th}}$ & {0.823} ~~~~ &{0.535} ~~~~ & {0.159} ~~~~ & ${0.689 \times 10^{-2}}$ \\ \hline 
$\Delta u_{\RPA}/\Delta u$ ~~~~~~& 0.97& 0.94& 0.94& 0.98 \\ 
$\Delta P_{\RPA}/\Delta P$     & 1.07& 0.96& 0.92& 0.94 \\ 
\hline \hline
\end{tabular}
\label{hight_uz}
\end{center} 
\vspace*{-0.7cm}
\end{table}

We first assess the validity of RPA at temperatures above $T_{th}$. 
Table~\ref{hight_uz} shows the ratio of RPA to simulation results of the
fluctuation parts of the potential energy and pressure at $T=T_{th}$. 
The deviations are smaller than 10\% and they monotonically
become smaller as the temperature increases, implying the
thermodynamic quantities are well-described by RPA even at $T_{th}$.  

\begin{figure}[t]
\begin{center}
\includegraphics[width=0.95\columnwidth]{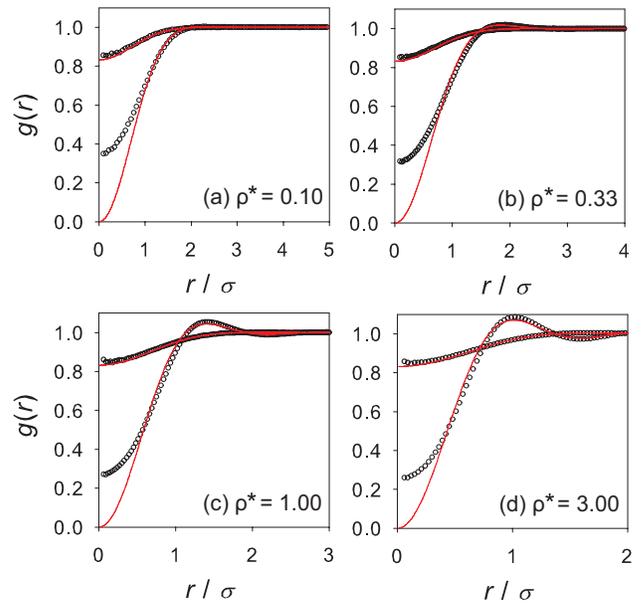}
\caption{
The radial distribution function in the high temperature regime for (a)
 $\rho^{\ast}=0.10$, (b) $0.33$, (c) $1.00$, and (d) $3.00$. 
Circles and lines are simulation and RPA results, respectively.
Two results in each panel correspond to the results at the threshold 
temperature ($T^* =$ 0.823, 0.535, 0.159, and $0.689 \times 10^{-2}$)
and higher temperature at which $g_{\RPA}(r=0) = 0.83$ ($T^* =$ 5.76, 5.35, 4.30, and 2.29). 
}
\vspace*{-0.3cm}
\label{hight_gofr}
\end{center}
\end{figure}

We also computed the radial distribution function $g(r)$.
Fig.~{\ref{hight_gofr}} shows $g(r)$ obtained from simulation (open
circles) and RPA (solid lines) at $T=T_{th}$ and at much higher
temperatures at which $g_{\RPA}(r=0) = 0.83$, for several densities. 
At high temperatures, agreement of simulation results with RPA is
excellent in all densities and for all $r$'s. 
At $T=T_{th}$, however, RPA works poorly around $r=0$, as expected
from Eq.~(\ref{gRPA}).
On the other hand, RPA's results perfectly match with the simulation results 
at larger $r$'s including the first shell peaks. 
The reason why thermodynamic quantities are well described by RPA even
at $T_{th}$ whereas agreement of $g(r)$ near $r=0$ is poor can be
attributed to the fact that the short-range part of $g(r)$ does not
contribute to both the potential energy and pressure, as 
we shall discuss in the following subsection.

\subsection{Low temperature regime}

We move to temperatures below $T_{th}$ and discuss the validity of
RPA in describing thermodynamic properties of GCM at high densities. 
In Fig.~\ref{sk}, the static structure factors obtained from RPA,
$S_{\RPA}(k)$, are shown in the solid lines.
It is obvious that RPA cannot capture even the qualitative behaviors
of  $S(k)$; 
$S_{\RPA}(k)$ remains flat at higher $k$'s and does not possess 
any prominent peak. 
However, as the main panel and inset of Fig.~\ref{sk} shows, 
RPA correctly predicts the low $k$'s behavior up to 
just below the wavevector at which the first peaks are located. 
It is in stark contrast with ordinary simple atomic fluids 
for which RPA works poorly for the whole range of wavevectors
\footnote{Even for the one component plasma,  
a typical long-range interacting system for which RPA is often employed
to describe its static properties, 
RPA does not capture the structures at low $k$'s. 
See Ref.\cite{Ichimaru1982}.}.
The excellent agreement implies that the mean-field character
of the dense GCM still survives in the length scales slightly longer than the
typical interparticle distance.  

Next, we compare RPA with simulation results for thermodynamic quantities.
We first look at the equation of state. 
In Fig.~\ref{prhot}, the pressure obtained from RPA are plotted in solid
lines. 
As shown in Fig.~\ref{prhot} (a) and (b), the deviation of the values of
RPA from those of simulation is very large at $\rho^{\ast}=0.33$
and the discrepancies increase with decreasing temperature. 
RPA also predicts a fictitious negative thermal expansion coefficient 
at this density as shown in Fig.~\ref{prhot} (b). 
In high densities, however, agreement of RPA with simulation 
is excellent for all temperatures down to the freezing temperatures
as shown in Fig.~\ref{prhot} (c)--(f). 
This is very surprising because the temperatures in these figures are
far below $T_{th}$, where RPA fails to describe overall shapes of 
$S(k)$ as discussed in the previous subsection.
 
We also calculated $T_{\alpha,\RPA}$ by solving $\partial
P_{\RPA}/\partial T = 0$ and plotted in Fig.~\ref{spd} (dotted line).  
Agreement of $T_{\alpha,\RPA}$ with the simulation result is perfect
except for the vicinity of the re-entrant melting region. 
The asymptotic expression of $T_{\alpha,\RPA}$ which is valid at high
densities can be written as (see Appendix) 
\begin{eqnarray} 
\frac{\kb T_{\alpha,\RPA}}{\epsilon} = \pi^{3/2} \rho\sigma^3 \exp\ \left[- \left( 
\frac{ {15} \pi^2}{4} \rho\sigma^3 \right)^{2/5} -2 \right]. 
\label{alpha}
\end{eqnarray} 
This asymptotic expression works well down to $\rho^{\ast} \sim 1$ as in the case of $T_{th}$. 
Interestingly, the density exponent $2/5$ in this expression is 
smaller than $2/3$ for $T_{th}$ and $T_m$ (see Eqs.~(\ref{st}) and (\ref{th})). 
Due to this difference, $T_{\alpha}$ monotonically deviates from $T_m$ as
the density increases and it eventually becomes larger than $T_{th}$ in
the high density limit (not shown).

\begin{figure}[t]
\begin{center}
\includegraphics[width=0.95\columnwidth]{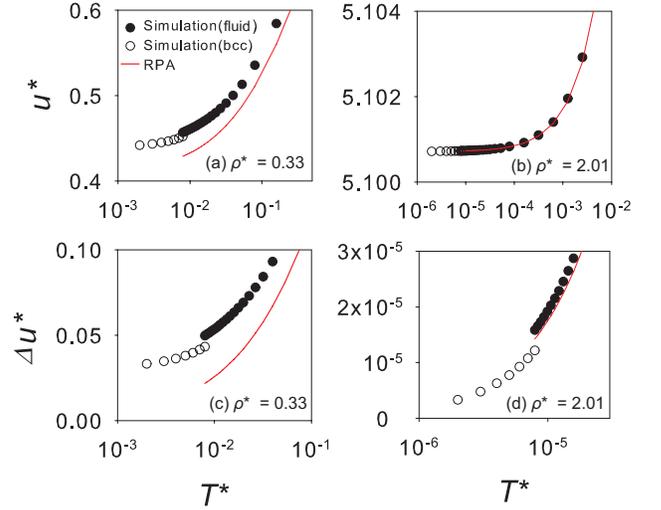}
\caption{
The temperature dependence of $u$ and $\Delta u$
at $\rho^{\ast}=0.33$ ((a) and (c)) and 
$\rho^{\ast}=2.01$ ((b) and (d)).
Filled and open circles represent the simulation results for the fluid
and crystalline (bcc)  phase, respectively.
Solid lines are the RPA results for the fluid phase. 
}
\vspace*{-0.3cm}
\label{ut}
\end{center}
\end{figure}

We made similar comparison for the potential energy in Fig.~\ref{ut}.
Fig.~\ref{ut} (a) and (b) show the temperature dependence of the
potential energy in the fluid phase (filed circles) and crystalline phase
(open circles) at two densities. 
The uniform part of $u$ is temperature independent (see
Eq.~(\ref{ueq})), although it dominates the net values of $u$. 
In order to see the temperature dependence of $u$ more clearly, the fluctuation
part $\Delta u$ are shown in Fig.~\ref{ut} (c) and (d). 
One observes that agreement of the simulation results with RPA is far better at high
densities;     
At $\rho^{\ast}=0.33$, RPA underestimates $\Delta u$ and
discrepancy from the simulation data are larger than the energy gap
between the fluid and crystalline phases. 
At the higher density $\rho^{\ast}=2.01$, the discrepancy is less 10\% even
at the freezing temperature. 

\begin{figure}[t]
\begin{center}
\includegraphics[width=0.95\columnwidth]{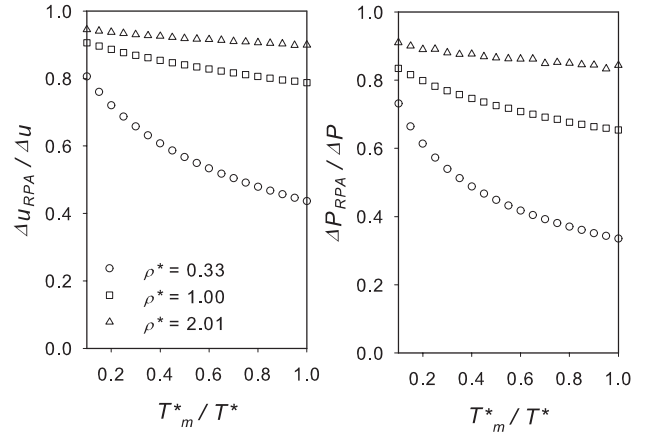}
\caption{
Temperature dependence of the ratio of RPA to simulation values of the 
fluctuation part of 
the potential energy (left) and the pressure (right) 
for $\rho^{\ast}=0.33$ ($\bigcirc$),
$1.00$ ($\square$), and $2.01$ ($\bigtriangleup$). 
The range of temperatures in these  figures is much lower than the
corresponding $T_{th}$.
($T_m^*/T_{th}^*$ are $1.48 \times 10^{-2}$, $3.14 \times 10^{-3}$, and $2.55 \times 10^{-4}$ 
for $\rho^{\ast}=0.33$, $1.00$, and $2.01$, respectively.) 
}
\vspace*{-0.3cm}
\label{lowt_uz}
\end{center}
\end{figure}

In order to quantify the accuracy of RPA for both the potential
energy and pressure, 
we plot the ratios of $\Delta u$ and $\Delta P$ of simulation to those of RPA
in Fig.~\ref{lowt_uz} against the inverse temperatures for several densities.  
The ratios decrease rapidly as the temperature decreases at the low density $\rho^{\ast}=0.33$,
whereas, in the high density $\rho^{\ast}=2.01$, the ratios for both the
potential energy and pressure remain to be more than 80\% for the whole
range of temperatures down to the melting temperature. 

\begin{figure}[t]
\begin{center}
\includegraphics[width=0.95\columnwidth]{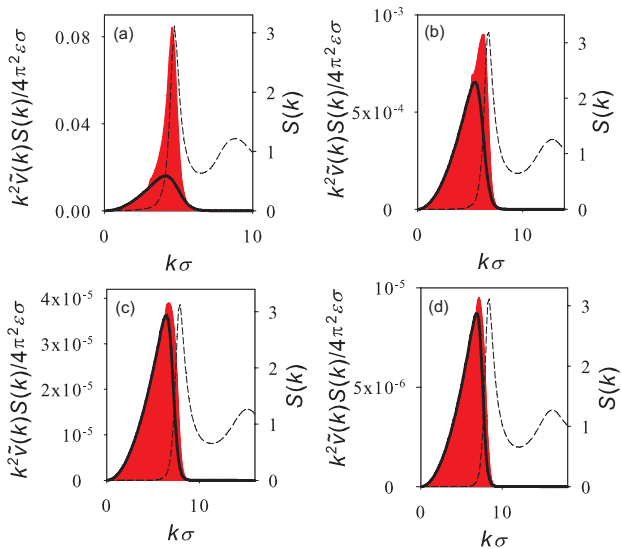}
\caption{
The integrand of Eq.~(\ref{du1}) (shaded areas), 
the integrand where $S(k)$ is replaced with $S_{\RPA}(k)$  (solid lines)
and $S(k)$ (broken lines) for 
(a) $T^{\ast} = 7.9 \times 10^{-3}$ and $\rho^{\ast}=0.33$, 
(b) $T^{\ast} = 5.0 \times 10^{-4}$ and $\rho^{\ast}=1.00$, 
(c) $T^{\ast} = 3.2 \times 10^{-5}$ and $\rho^{\ast}=1.65$ and 
(d) $T^{\ast} = 7.9 \times 10^{-6}$ and $\rho^{\ast}=2.01$. 
}
\vspace*{-0.3cm}
\label{lowt_int}
\end{center}
\end{figure}

Given that RPA poorly describes even the qualitative behaviors of $S(k)$ and $g(r)$
at the high density and low temperature regimes,  it is very surprising
and counterintuitive that RPA is excellent at predicting quantitatively 
thermodynamic quantities $u$  and $P$. 
In order to rationalize this puzzling facts, 
we look again the integral expressions of the thermodynamic functions,
Eq.~(\ref{du1}) and (\ref{dz}). 
These expressions show that both $\Delta u$ and $\Delta P$ are
expressed in terms of the integral of $S(k)$ multiplied with the pair
potential over the wavevectors. 
In order to see which length scales dominate the integrand, 
we show the integrand of $\Delta u$ in Eq.~(\ref{du1}) with simulated $S(k)$
in Fig.~\ref{lowt_int} (shaded area), 
along with those obtained using $S_{\RPA}(k)$ (solid line).
At the low density $\rho^{\ast}=0.33$, the integrand with 
simulated $S(k)$ is peaked at the 
peak position of $S(k)$ (broken line). 
This implies that $\Delta u$ at this density 
is dominated by the contribution at the
interparticle distance, just like ordinary atomic fluids.  
The integrand obtained using RPA fails to account for this peak
structure (solid line). 
However, with increasing density, the peak position of the integrand
shifts to smaller wavevectors than the first peak position of $S(k)$. 
Concomitantly, agreement of the integrand obtained from simulation and RPA 
becomes better and better. 
This agreement originates from the fact that RPA can account excellently
for the low wavevector behavior of the static structure factor 
as shown in Fig.~\ref{sk}. 
At very high densities, the particles start overlapping and 
the characteristic interparticle distance decouples with the length
scales which dominate thermodynamic quantities of GCM. 
This is the reason why RPA remains the excellent approximation 
to predict thermodynamic quantities even far below the threshold temperatures. 

\section{Conclusions}

In this paper, we have presented a detailed analysis of thermodynamic
and structural properties of the high density one-component GCM. 
Special emphasis has been put for static properties of the fluid phase. 
First, the solid-fluid phase boundary of the system is carefully
evaluated up to the unprecedentedly high density $\rho^{\ast} =2.4$. 
Our result confirmed the scaling conjectured by Stillinger for the
freezing and melting temperatures, 
$\log T_f, ~\log T_m \propto -\rho^{2/3}$, at $\rho^\ast \gtrsim 1.2$. 
The potential energy difference between the crystalline and fluid phases was
shown to be linear in the freezing temperature and 
the entropy difference is almost constant at high densities, which
verifies the assumption which Stillinger's argument is based upon.
The thermodynamic and structural properties of GCM in the fluid phase
are analyzed in detail for a wide range of temperature and density. 
The potential energy $u$, the equation of state $P$, the static structure
factor $S(k)$, and the radial distribution function $g(r)$ were evaluated
by simulation.
We compare the simulation results with the RPA results.
In the high temperature regime, RPA provides almost perfect description for
both thermodynamic quantities and the structural factor $S(k)$.   
RPA is rather poor at predicting $g(r)$ at $r \approx 0$.
Threshold temperature $T_{th}$ below which RPA fails to describe $g(r=0)$
is relatively high. 
In the high density and low temperature regime, RPA fails to capture the
peak structure of $S(k)$ even qualitatively, whereas 
it predicts correctly the low $k$'s behavior up to 
just below the wavevector at which the first peaks are located. 
Despite of poor performance of RPA at describing the structural
properties, 
RPA successfully describes thermodynamic quantities such as 
the potential energy and pressure at high densities. 
Agreement of RPA with simulation results systematically improves as the
density increases even near the phase boundary. 
The temperature below which the thermal expansion coefficient become
negative is also accurately calculated from RPA.  
By scrutinizing the role of the microscopic structure of particles 
in the potential energy and pressure, we concluded that the
surprising success of RPA is originated from the decoupling of the
length scales which dictate the thermodynamic quantities 
and the interparticle distance. 
This decoupling is attributed to the mild and long-ranged repulsive tails of
the pair potential of GCM. 
The fact that RPA is an excellent approximation even at the vicinity of
the phase boundary, or even at the supercooled regime, at high
densities, hints that the mean-field description is valid for the high
density GCM and may play a crucial role to understand (glassy)
dynamics let alone thermodynamic properties~\cite{Ikeda2011,Ikeda_II}.

\acknowledgments
This work is partially supported by Grant-in-Aid for JSPS Fellows (AI), KAKENHI; \# 21540416, (KM), and Priority Areas ``Soft Matter Physics'' (KM). 

\appendix
\section{Derivation of Eqs.~(\ref{th}) and (\ref{alpha})}

In this appendix, we derive the asymptotic expressions of $T_{th}$ and $T_{\alpha,\RPA}$ 
at high densities 
by using the asymptotic expansion of the polylogarithm~\cite{Wood1992} 
\begin{eqnarray}
\Li_{\nu}(-x) = - \frac{(\log x)^{\nu}}{\Gamma(\nu + 1)} + \mathcal{O}((-\log x)^{\nu - 2}), \label{polylog}
\end{eqnarray}
where $\Gamma(x)$ is the gamma function. 

\begin{figure}[t]
\begin{center}
\includegraphics[width=0.95\columnwidth]{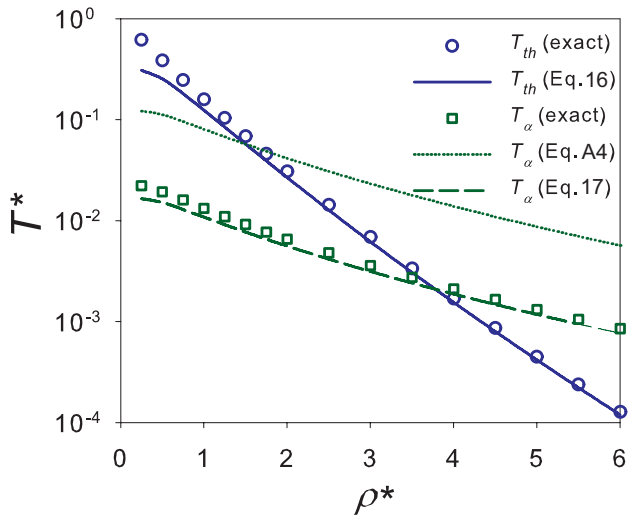}
\caption{
The validity of the asymptotic expressions for $T_{th}$ and $T_{\alpha,\RPA}$. 
Open circles and solid line are RPA values and its asymptotic expression, Eq.~(\ref{th}), of $T_{th}$, respectively. 
Open squares, dotted line, and dashed line are RPA values and its two
 asymptotic expressions, Eqs.~(\ref{eq0}) and (\ref{alpha}), of $T_{\alpha,\RPA}$, respectively. 
}
\vspace*{-0.3cm}
\label{asymptotic}
\end{center}
\end{figure}
At $T=T_{th}$, Eqs.~(\ref{gRPA}) and (\ref{polylog}) lead to
\begin{equation}
 \begin{aligned}
&
\rho^\ast - \frac{4}{3 \pi^2} 
\left(\log x_{th} \right)^{3/2}
+ \mathcal{O}\left(\left(\log x_{th}\right)^{-1/2}\right) = 0, 
 \end{aligned}
\end{equation}
where $x_{th}= \pi^{3/2} \rho^\ast/T_{th}^{\ast}$.
When $\rho^\ast$ is sufficiently large, we can neglect the third term
on the left hand side, 
leading to the asymptotic expression of $T_{th}$ of Eq.~(\ref{th}). 
Fig.~\ref{asymptotic} shows that this asymptotic expression is very
accurate down to $\rho^{\ast} \sim 1$. 
Likewise, at $T=T_{\alpha,\RPA}$, Eqs.~(\ref{upRPA}) and (\ref{polylog}) lead to 
\begin{equation}
 \begin{aligned}
&
\rho^\ast 
- \frac{4}{15 \pi^2} \left(\log x_{\RPA}\right)^{5/2}  
+ \frac{4}{3 \pi^2}  \left(\log x_{\RPA}\right)^{3/2}
\\
& 
+ \mathcal{O}\left(\left(\log x_{\RPA}\right)^{1/2}\right) = 0, 
  \end{aligned}
\label{ap1}
\end{equation}
where $x_{\RPA} = \pi^{3/2} \rho^\ast/T_{\alpha,\RPA}^\ast$.
If we neglect the third and fourth terms on the left hand side, we  obtain  
\begin{eqnarray}
T_{\alpha,\RPA}^{\ast}
 = \pi^{3/2} \rho^\ast \exp\left[-\left(\frac{15\pi^2\rho^\ast}{4}\right)^{2/5}\right]. \label{eq0}
\end{eqnarray}
However, Fig.~\ref{asymptotic} shows that this expression is not accurate even at very high densities. 
For sufficient accuracy, we have to keep the third term on the left hand side of Eq.~(\ref{ap1}). 
If we neglect only the forth term on the left hand side, Eq.~(\ref{ap1}) becomes
\begin{eqnarray}
\frac{4}{15 \pi^2} s^{5/2} - \rho^{\ast -2/5} \frac{4}{3 \pi^2} s^{3/2} - 1 = 0, 
\end{eqnarray}
where $s = \rho^{\ast -2/5} \log (\pi^{3/2} \rho^\ast /T_{\alpha}^\ast)$.
Since the second term on the left hand side of this equation is small
due to the factor $\rho^{\ast -2/5}$, 
we can expand the solution as $s = s_0 + \rho^{\ast -2/5} s_1 +
\cdots$ and can solve the equation in each order of density.  
The first order solution including $s_0$ and $s_1$ is 
\begin{eqnarray}
s = \Bigl( \frac{15 \pi^2}{4} \Bigr)^{2/5} + 2 \rho^{\ast -2/5},  
\end{eqnarray}
which leads to the asymptotic expression of $T_{\alpha,\RPA}$ of Eq.~(\ref{alpha}).  
Fig.~\ref{asymptotic} shows that this expression is very accurate down to  $\rho^{\ast} \sim 1$.

\end{document}